\documentclass[10pt,conference]{IEEEtran}
\IEEEoverridecommandlockouts
\usepackage{cite}
\usepackage{amsmath,amssymb,amsfonts}
\usepackage{algorithmic}
\usepackage{graphicx}
\usepackage{textcomp}
\usepackage{subcaption}
\usepackage{tikz}
\usepackage{xcolor}
\usetikzlibrary{positioning}
\def\BibTeX{{\rm B\kern-.05em{\sc i\kern-.025em b}\kern-.08em
    T\kern-.1667em\lower.7ex\hbox{E}\kern-.125emX}}
\begin{document}

\title{Predicting Expressibility of Parameterized Quantum Circuits using Graph Neural Network

\thanks{\small{Research presented in this article was supported by the Laboratory Directed Research and Development program of Los Alamos National Laboratory under project number 20230049DR.}}
}

\author{
    \IEEEauthorblockN{
    Shamminuj Aktar\IEEEauthorrefmark{1}, 
    Andreas Bärtschi\IEEEauthorrefmark{2}, 
    Abdel-Hameed A. Badawy\IEEEauthorrefmark{1}, 
    Diane Oyen\IEEEauthorrefmark{2} and 
    Stephan Eidenbenz\IEEEauthorrefmark{2}}
    \IEEEauthorblockA{\IEEEauthorrefmark{1}
    \textit{Klipsch School of Electrical and Computer Engineering, New Mexico State University},
    Las Cruces, NM, 88001, USA \\
    Email: saktar@nmsu.edu,badawy@nmsu.edu
    }
    \IEEEauthorblockA{\IEEEauthorrefmark{2}
    \textit{CCS-3 Information Sciences, Los Alamos National Laboratory},
    Los Alamos, NM 87544, USA \\
    Email: baertschi@lanl.gov, doyen@lanl.gov, eidenben@lanl.gov
    }
}
\maketitle

\begin{abstract}

    Parameterized Quantum Circuits (PQCs) are essential to quantum machine learning and optimization algorithms. The expressibility of PQCs, which measures their ability to represent a wide range of quantum states, is a critical factor influencing their efficacy in solving quantum problems. However, the existing technique for computing expressibility relies on statistically estimating it through classical simulations, which requires many samples. 
    In this work, we propose a novel method based on Graph Neural Networks (GNNs) for predicting the expressibility of PQCs. By leveraging the graph-based representation of PQCs, our GNN-based model captures intricate relationships between circuit parameters and their resulting expressibility. We train the GNN model on a comprehensive dataset of PQCs annotated with their expressibility values. Experimental evaluation on a four thousand random PQC dataset and IBM Qiskit's hardware efficient ansatz sets demonstrates the superior performance of our approach, achieving a root mean square error (RMSE) of 0.03 and 0.06, respectively. 
    %

\end{abstract}

\vspace{-1ex}
\section{Introduction}
    Parameterized Quantum Circuits (PQCs) are essential in leveraging the capabilities of quantum computers. They provide a flexible framework for solving intricate problems by optimizing tunable parameters within a quantum circuit. A PQC is a sequence of gates applied to a set of qubits, with some gates allowing for adjustable classical parameters that can be varied during circuit execution to prepare a quantum state. The importance of PQCs has led to the development of new ansatz designs, e.g.,~problem-specific PQCs and hardware-efficient PQCs. Researchers have proposed different qualitative metrics i.e. expressibility~\cite{sim2019expressibility}, entangling capability~\cite{sim2019expressibility}, and trainability~\cite{mcclean2018barren} to estimate the quality and usefulness of PQCs.

    \begin{figure}
      \centering
      \includegraphics[width=0.98\linewidth]{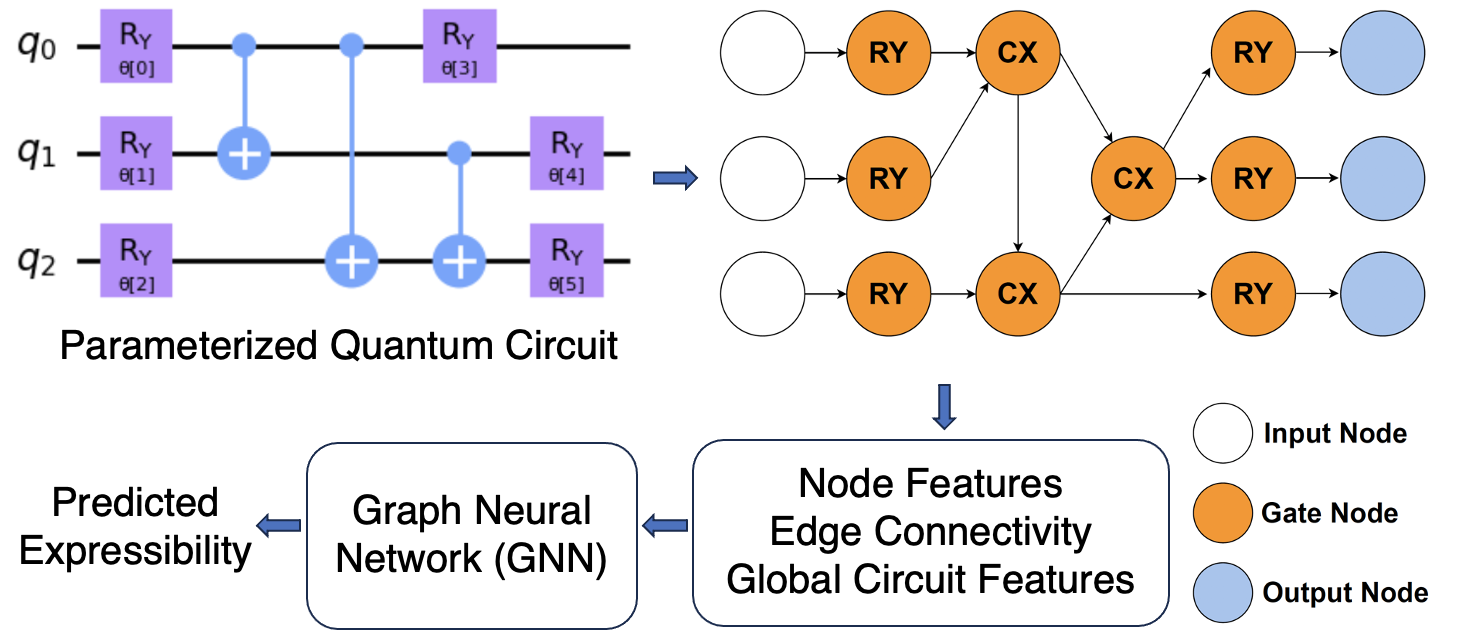}
      \caption{Our proposed framework for predicting expressibility of PQC. We derive a graph representation of PQC with input, gate, and output nodes. Nodes are encoded with feature vectors representing node type and utilized qubits, edges capture information flow within the PQC, and global features encompass circuit characteristics. 
      }
      \label{fig:diagram}
      \vspace{-2ex}
    \end{figure}

    The expressibility of a PQC~\cite{sim2019expressibility} determines its ability to explore states in the Hilbert space. Expressibility is measured by computing the Kullback-Leibler (KL) divergence~\cite{joyce2011kullback} between the estimated fidelity distribution $P_{PQC}(F;\theta)$ measured as a distance between sampled pairs of parameterized quantum states, and the maximally expressive uniform distribution resulting from a Haar-random unitary $P_{Haar}(F)$~\cite{zyczkowski2005average}. 
    \begin{equation}
        \label{eq:expr}
        Expr = D_{KL} (P_{PQC}(F;\theta) || P_{Haar}(F))
    \end{equation}  
    Lower KL divergence indicates better expressibility, meaning a better ability to explore a wider range of states in the Hilbert space. The fidelity distribution required in expressibility computations measures the overlap between two quantum states prepared using different sets of parameters. Measuring expressibility directly requires many samples, which are time-consuming to acquire. For instance, Sim et al. employed 5000 fidelity samples for four-qubit circuits to measure expressibility. This reliance on a large number of samples poses challenges due to the time and computational resources required. 

    To address these challenges, we propose a novel approach utilizing a Graph Neural Network (GNN)~
    \cite{scarselli2008graph} to predict the expressibility of PQCs shown in Figure~\ref{fig:diagram}. GNNs have demonstrated exceptional capabilities in capturing complex relationships within graph-structured data, making them well-suited for analyzing the intricate structure of PQCs. We train and evaluate our GNN model on a comprehensive dataset of PQCs to demonstrate the effectiveness of our approach. Evaluation on around four thousand random PQCs and IBM Qiskit's hardware efficient ansatz sets demonstrate the high accuracy of the model, RMSE of 0.03 and 0.06, respectively. 
\begin{figure*}
  \centering
  \begin{subfigure}{0.375\linewidth}
    \centering
    \includegraphics[width=\linewidth]{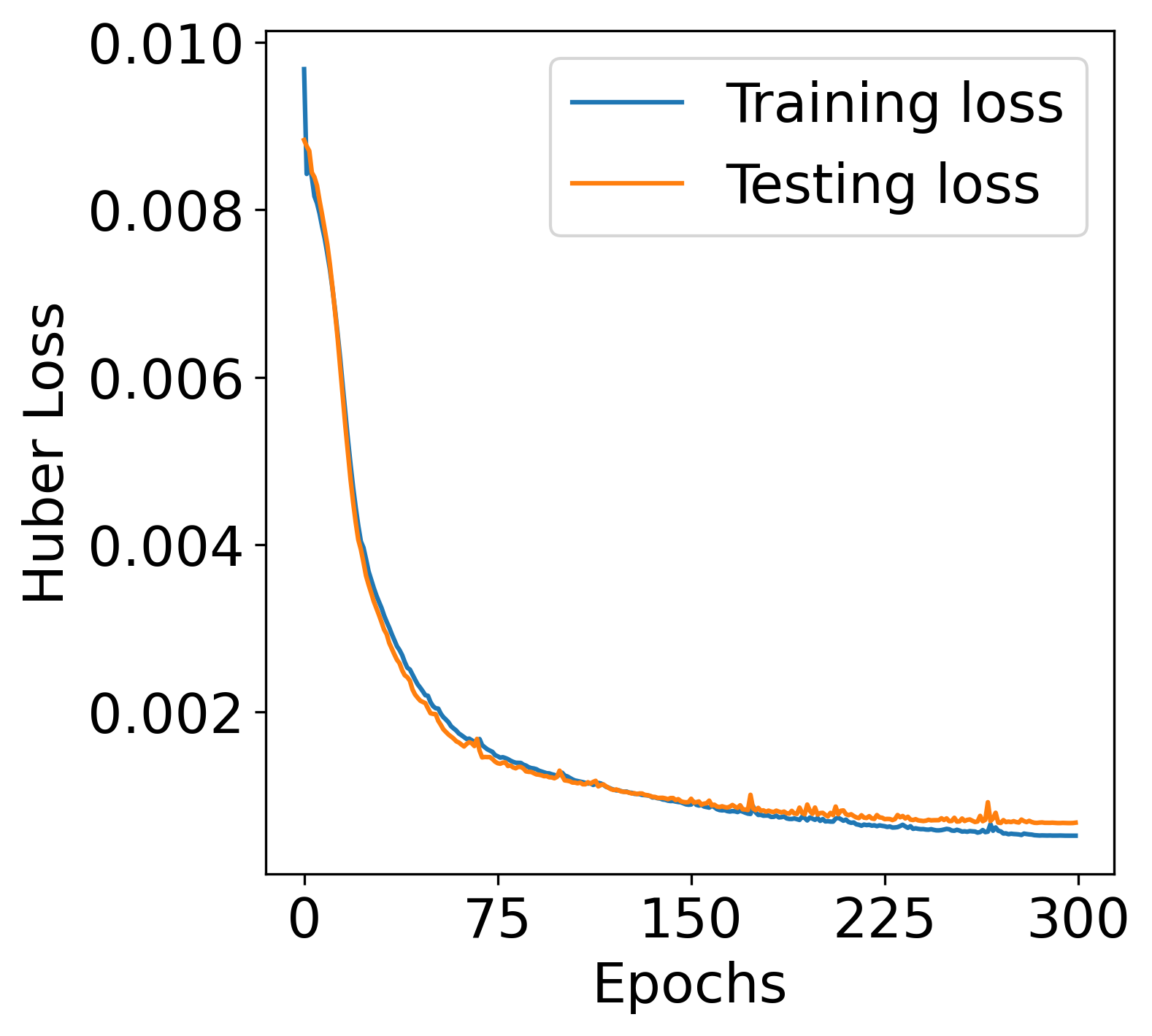}
  \end{subfigure}
  \begin{subfigure}{0.34\linewidth}
    \centering
    \includegraphics[width=\linewidth]{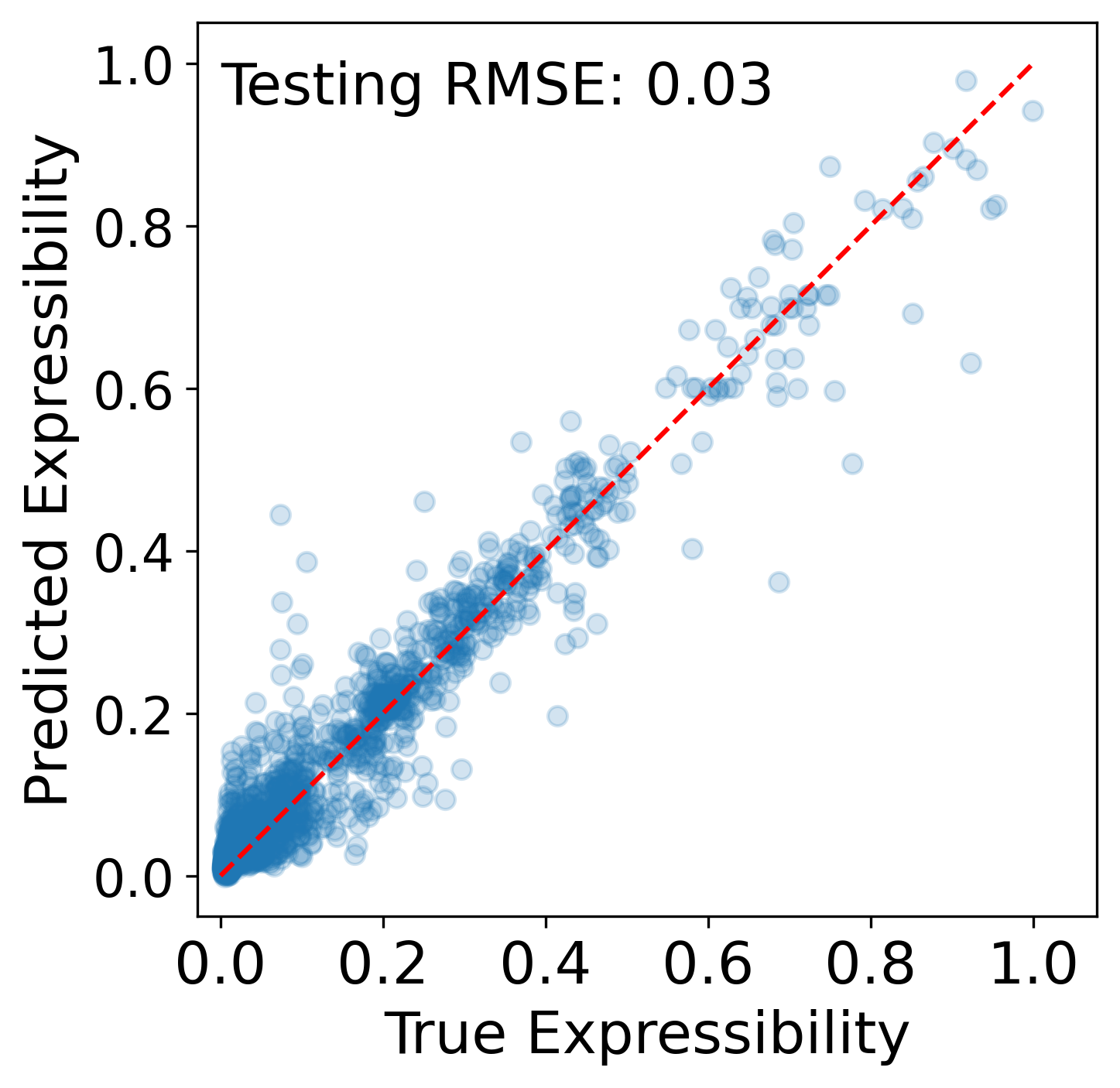}
  \end{subfigure}
  \caption{\emph{(Left)}: Prediction loss of training and testing datasets with the number of epochs during GNN model training shows the generalizability of the model. \emph{(Right)}: Trained model's predicted expressibility compared to the true expressibility for the testing dataset yielded an overall RMSE of 0.03}
  \label{fig:combined}
  \vspace{-2ex}
\end{figure*}
\section{Proposed Framework}
\subsection{Dataset Generation}
In our study, we generate random parameterized quantum circuits using a qubit gate set of X, SX, RX, RY, RZ, and CX gates. The first layer consists of random single-qubit gates for each qubit. That layer is followed by random entanglement between CX gates. The entanglement was established by selecting two random qubits and applying the CX gate between them. The length of the entanglement, representing the number of consecutive CX gates, was also chosen randomly. We then repeated the layers of single-qubit gates and CX entanglement multiple times to deepen the circuit. Finally, we concluded the circuit with another layer of single-qubit gates. Overall, we generated 12,000 circuits, with a maximum of 4 qubits and a maximum depth of 40. Next, we compute the estimated fidelity distribution for each circuit using the quantum kernel method to compute the distance between pairs of parameterized quantum states. We run the experiments in IBM QASM simulator and then we compute the expressibility values of the circuits using Equation~\ref{eq:expr}.
Additionally, we include 64 hardware-efficient ansatz circuits from Qikit's RealAmplitude circuits up to four qubits with up to 4 circuit repetitions with different entanglement patterns available as validation dataset.
\subsection{Graph Neural Network (GNN) Model}
We utilize a GNN model to learn the complex relationship between PQCs and their expressibility. First, we extract a directed graph with a set of nodes, i.e., input, output, and gate nodes, and edge connectivity showing the flow of information within the PQC. Each node has a one-hot encoded vector representing features like node type and qubits. The circuits also include global feature sets, i.e., circuit depth, width, number of parameterized gates, number of qubits, and counts of different gates represented as a vector and fed to the GNN model using three fully connected (FC) layers. First, our architecture employs three SAGEConv layers to capture and process the local neighborhood information within the PQC graphs. Next, the global feature vector is concatenated with the aggregated node feature vector, and the combined representation is fed into the model. Finally, there is a regression layer to predict the expressibility value of the PQC.

\begin{figure}
  \centering
  \includegraphics[width=0.95\linewidth]{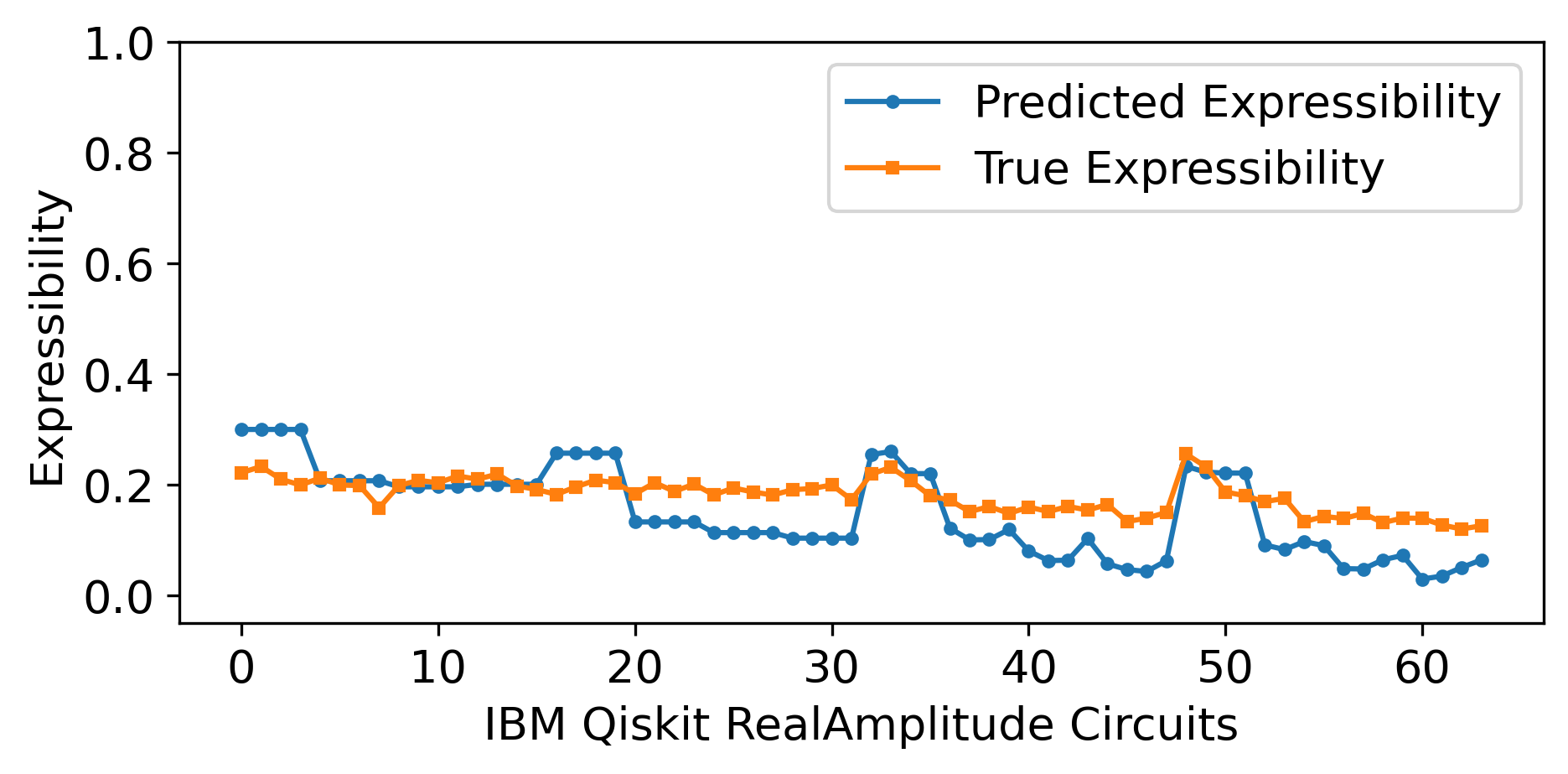}
  \caption{Predicted expressibility compared to true expressibility for the 64 Qiskit's RealAmplitude circuits resulted in an overall RMSE of 0.06.}
  \label{fig:realamplitude-rmse}
  \vspace{-3ex}
\end{figure}

\subsection{Training and Evaluation}
We use 80\% of the randomly generated PQC dataset to train our GNN model. Before training, we normalize the node features across the dataset by removing the mean and dividing the standard deviation. We train the model using Adam optimizer for 300 epochs with learning rate $10^{-4}$ and weight decay $10^{-6}$ and ReduceLROnPlateau scheduler that reduces the learning rate by a factor of 0.1. We utilize a batch size of 2048, employ the Huber loss as our loss function and save the model with the best validation loss. Figure~\ref{fig:combined}(left) shows the training and testing prediction loss as the number of epochs increases during training.
Our trained model achieved an RMSE of 0.03 for our testing dataset consisting of 20\% randomly generated PQCs. Figure~\ref{fig:combined}(right) shows scatter plots of model predicted expressibility vs. the true expressibility for our test dataset. Additionally, we evaluated our model using 64 circuits from IBM Qiskit's RealAmplitude circuits and achieved an RMSE of 0.06. Figure~\ref{fig:realamplitude-rmse} demonstrate predicted vs. true expressibility for 64 RealAmplitude circuits. These results demonstrate the effectiveness of the GNN model in predicting the expressibility of PQCs.

\section{Conclusion}
We present a GNN-based approach for predicting the expressibility of PQCs, considering the PQC as a graph. By leveraging the capabilities of GNNs, we effectively capture the intricate relationships between the structure of PQCs and their expressibility. Evaluation on the random PQC dataset and Qiskit's RealAmplitude circuit sets shows effective accuracy of our expressibility prediction technique. Importantly, this approach significantly reduces the cost associated with computing fidelity distribution for a large number of samples in expressibility computation.
\bibliographystyle{IEEEtranS}
\bibliography{main}
\end{document}